\documentclass[twocolumn]{autart}    

\usepackage{graphicx}    
\usepackage{times} 
\usepackage{amsmath,amssymb}
\usepackage{setspace}
\usepackage{algorithm}
\usepackage{algpseudocode} 
\usepackage{url} 
\usepackage{color}
 \usepackage{float}
 
\newtheorem{Theorem}{Theorem}
\newtheorem{Proposition}{Proposition}

\newtheorem{Porblem}{Porblem}
\newtheorem{Definition}{Definition}
\newtheorem{Example}{Example}

\begin{document}

\begin{frontmatter}

\title{To Transmit or Not to Transmit: 
Optimal Sensor Schedule for 
Remote State Estimation of Discrete-Event Systems\thanksref{footnoteinfo}} 

\thanks[footnoteinfo]{This paper was presented at ACC2022. Corresponding author: Jin Hu.}

\author[a]{Yingying Liu}\ead{yingyingliu611@163.com},    
\author[a]{Jin Hu}\ead{hujin007@nwafu.edu.cn},               
\author[a]{Yongxia Yang}\ead{yangyongxia@nwafu.edu.cn},  
\author[Wei]{Wei Duan}\ead{dwei1024@126.com}

\address[a]{School of Mechanical and Electronic Engineering, Northwest A $\&$ F University, Xi'an, China} 
\address[Wei]{School of Electro-Mechanical Engineering, Xidian University, Xi'an, China}             

\begin{keyword}                           
State estimation; Discrete-event systems; Information transmission policy.               
\end{keyword}                             

\begin{abstract}                          
This paper considers the problem of optimal sensor schedules for remote state estimation of discrete-event systems. 
In this setting, the sensors observe information from the plant and transmit the observable information to the receiver or estimator selectively. A transmission mechanism decides whether the observable information is transmitted or not, according to an {\it information transmission policy}, such that the receiver has sufficient information to satisfy the purpose of decision-making. 
To construct such a transmission mechanism, 
we first construct a non-deterministic dynamic observer that  contains all feasible information transmission policies. 
Then, we show that  the information updating  rule  of  the  dynamic  observer indeed yields the state estimate from  the receiver's point of view. 
Finally, we propose an approach to  extract  a specific information transmission policy, realized by a finite-state automaton, from the dynamic observer while satisfying some desired observation properties. 
To reduce transmission-related costs, we also require that the sensors transmit events as few as possible. 
A running example is provided to illustrate the proposed procedures.
\end{abstract}

\end{frontmatter}
\section{Introduction}
Partially-observed discrete-event systems (DES) \cite{zhou2022state,ma2023verification} have attracted widespread attentions because, in the real physical environment, some events are unobservable due to the limited observation cost or observation capacity. Examples include automated guided vehicles \cite{ElMaraghy2005}, manufacturing machines \cite{hu2023design,liu2019,PAIVA2021100146}, and robots \cite{ZHOU2020,YinRobust}. State estimation \cite{han2022revisiting,li2022synthesis} is increasingly important in the analysis of partially-observed DES for the purposes of decision-making \cite{ALIKHANI2021109575} or system diagnosis \cite{li2023error,yang2022secure}. 

 \begin{figure}
  \centering
  \includegraphics[width=0.46\textwidth]{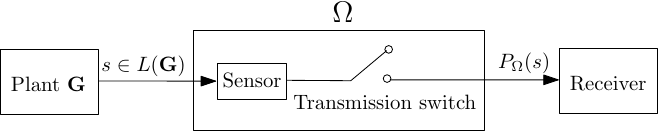}\\
  \caption{Architecture of transmission mechanism, where $L(G)$ denotes generated behavior of the plant $G$ and $\Omega$ denotes the transmission policy and $P_{\Omega}$ denotes the information mapping under policy $\Omega$ (see  precise definitions in Section~II).}
  \label{fig.1}
\end{figure} 
In this paper, we investigate \textit{remote} state estimation problem of partially-observed DES, which is widely used in the scenario that  the sensor and the consumer of the information are physically different and located at a remote distance \cite{alves2023state,liu2021improved,han2022revisiting,hou2023modeling}.  
As shown in Fig.~\ref{fig.1}, the system makes observations online through its sensors who read and sent the observed information to the consumer/receiver. The sensors can be turned ON/OFF dynamically by a transmission switch, where the switch decides whether to transmit its observed information to a receiver or not, according to \emph{an information transmission policy}.  
It is precisely because of such architecture 
that we do not need to  transmit an observed information all the time and only need to transmit it when it is necessary (to satisfy some observational property), which effectively reduces  sensor-related costs, e.g., bandwidth, energy, or security constraints.  
Then, the receiver makes control decisions for the system based on the transmitted information. 

Instead of investigating the enforcement of specific objectives, e.g., control or diagnosis, in this paper, we consider a particular class of properties called Information-State-based (IS-based) properties \cite{yin2016uniform} (see strict definitions in Section~\ref{pf}), where IS-based properties can be expressed in terms of suitably-defined information states. 
Roughly speaking, an IS-based property is a property that only depends on the current local information of the system.
Our objective is to synthesize an information transmission policy via the transmission mechanism shown in Fig. \ref{fig.1} such that some of given IS-based properties hold.

To do so, we first synthesize a so-called \emph{dynamic observer} in a non-deterministic manner that enforces IS-based property and contains all feasible information transmission policies. 
It is ensured that the synthesis problem of a feasible information transmission policy is always solvable. 
We show that the information updating rule of the dynamic observer indeed yields the state estimate of the receiver.
We then propose a method to extract a specific information transmission policy from the dynamic observer so as to ensure the given IS-based properties. 
Moreover, to reduce sensor-related costs, we require that the sensors transmit as few events as possible. 

Our work is closely related to an information acquisition mechanism  in \cite{yin2019synthesis}, called dynamic masks, that acquires information from the system by dynamically turning ON/OFF the associated sensors. 
The dynamic masks that preserve infinite-step opacity while maximizing the information acquired in \cite{yin2019synthesis}. A type of information state is proposed to capture all delayed information for all previous instants. By contrast, in this paper we give a new type of information state to analyse the system by synthesizing a dynamic observer. Moreover, in this paper we do not only consider the maximal solutions, but also investigate minimized solution in the sensor activation problem.  In addition, instead of consider the infinite-step opacity property, in this paper we consider a particular class of properties which contains the opacity property proposed in \cite{yin2019synthesis}.

The authors in \cite{yin2018minimization} propose a novel approach to find a language-based minimal sensor activation policy for each agent such that the agents can always make a correct global decision as a team. However, in \cite{yin2018minimization} the sensors are also considered as consumers of the information. The case that the sensor and the consumers are physically different is not considered. By contrast, this paper considers the remote estimation problem where the sensor and the consumers are located at a remote distance and thus be more complex and challenging. Also, authors in \cite{zhang2015maximum} investigate how to release the maximum information to the public by a controller while ensuring opacity, where the information is controlled by the controller. By contrast, in this paper, we consider the IS-based property, which is not restricted to opacity and can formulate safety,  diagnosability, detectability, and distinguishability. A communication scheme is developed for distributed agents in \cite{rudie2003minimal}, where each agent must be able to distinguish between some of its states to perform some control or monitoring task. In this work, we also consider the communication problem. However, instead of considering the communication between the agents, we consider the communication problem between the sensors and the receiver. 
Our aim is to design an information transmission policy  such that the sensors transmit as little information as possible.

In contrast to the conference version \cite{liu2022}, this paper contributes (i) a proof of the feasibility of the proposed policy; (ii) an algorithm for synthesizing a deterministic automaton from the largest feasible automaton, thus generating a specific information transmission policy; and (iii) an algorithm for illustrating the information transmission process at each step.


This paper is organized as follows. Section 2 presents preliminaries, critical definitions for subsequent sections, and the formulation of the information transmission problem. Section 3 outlines an approach for developing the structure of a dynamic observer that can realize all information transmission policies. Section 4 proposes a method for synthesizing a specific information transmission policy represented by an automaton. Finally, in Section 5, a brief summary is given for the paper.


\section{Preliminaries and Problem Formulation} \label{sec2_probm}
\subsection{System Model}
We consider a DES modeled by a deterministic finite-state automaton (DFA)
$\textbf{G}=(Q,\Sigma,\delta,q_0),$
 where  
 $Q$ is a finite set of states, 
 $\Sigma$ is a finite set of events, 
 $\delta\colon Q\times \Sigma\to Q$ is a (partial) transition function,  
 $q_0\in Q$ is the initial state. 
 In the usual way, $\delta$ can be extended to $\delta\colon Q\times \Sigma^{*}\to Q$, where $\Sigma^{*}$ is the set of all finite-length {\it strings}, including the empty string $\varepsilon$. In the case that $\delta\colon Q\times \Sigma^{*}\to 2^Q$, $\textbf{G}$ is a nondeterministic finite-state automaton (NFA).  The {\it generated behavior} of $\bf{G}$ is language $L(\textbf{G})=\{s\in \Sigma^{*}: \delta(q_0,s)!\}$, where $\delta(q_0,s)!$ means that $\delta(q_0,s)$ is defined. We say that a state $q\in Q$ is \emph{reachable} if there is a string $s\in\Sigma^*$ with $\delta(q_0,s)!$ and $\delta(q_0,s)=q$. 
We denote the set of events that are defined at state $q\in Q$ by $\Sigma_q=\{\sigma\in \Sigma:\delta(q,\sigma)!\}$.

A string $s_1\in \Sigma^{*}$ is a {\it prefix\/} of $\textit{s}\in \Sigma^{*}$, written as $s_1\leq s$, if there is a string $s_2\in \Sigma^{*}$ such that $s_1s_2 = s$. The length of a string $s$ is denoted by $|s|$. The {\it prefix closure\/} of a language $L$ is the set $\overline{L} = \{s\in\Sigma^{*} :\exists t\in L \text{ s.t. } s\leq t\}$. For a natural number $n$, let $[1,n]=\{1,\ldots,n\}$ denote the set of all natural numbers from 1 to $n$. Let $\Sigma_o \subseteq\Sigma$, we denote by $P:\Sigma^*\to \Sigma^*_o$ the standard natural projection from $\Sigma$ to $\Sigma_o$.

\subsection{Problem Formulation}\label{pf}
We consider the scenario where the observability properties of events can be controlled by an \emph{information transmission policy} during the evolution of the system. For the sake of simplicity, in this paper we assume that $\Sigma=\Sigma_o$, where $\Sigma_o$ is the set of events whose occurrences can always be observed by some sensors (full observation). 
An information transmission policy is defined as a labeling function   
\[
\Omega: \Sigma^*\Sigma\rightarrow \{Y,N\}
\]
that specifies the transmitted and non-transmitted events within $\Sigma$, where $Y$ and $N$ are \emph{information transmission labels}. Specifically, for each observation $s\sigma\in\Sigma^*\Sigma$,  $\Omega(s\sigma)=Y$ denotes the event $\sigma$ is transmitted after the occurrence of s; $\Omega(s\sigma)=N$ represents the opposite. While an event is transmitted, its current occurrence will be observed.  In other words, after string s, event $\sigma$ is currently ``unobservable" if $\Omega(s\sigma)=N$. The above definition of the information transmission policy is history-dependent, so the observability of an event with different histories may be different.

In practice, the  information transmission policy needs to be implemented in finite memory, which can be represented as a pair (a finite transducer)
\begin{align*}
\Omega=({\bf A},\mathcal{L}),
\end{align*}
where ${\bf A}=(X,\Sigma,\eta,x_{0})$ is a DFA, called a {\em sensor automaton}, such that i) $L({\bf A})=\Sigma^*$; ii) $ \forall x\in X,  \sigma\in \Sigma\setminus \Sigma: \eta (x,\sigma)=x$ (it means that unobservable events in $\Sigma$ are represented by slefloops in ${\bf A}$), and
$\mathcal{L}: X \times \Sigma\rightarrow  \{Y, N\}$ is a labelling function that specifies the transmitted events within $\Sigma$. 
For any $\sigma\in  \Sigma$, $\mathcal{L}(x,\sigma)=Y$ means that the occurrence of event $\sigma$ is transmitted at state $x$, while  $\mathcal{L}(x,\sigma)=N$ represents the opposite. $\mathcal{L}$ can be extended to $\mathcal{L}: X \times \Sigma^*\rightarrow  \{Y, N\}^*$. 

Note that we assume the event domain of ${\bf A}$ is $\Sigma$ for the sake of simplicity, but it can only update its sensor state upon the occurrences of its observable events in $\Sigma$. 
Hereafter in the paper, the information transmission policy will be considered as a pair $\Omega=({\bf A}, \mathcal{L})$ rather than a language-based mapping and let $\Sigma=\Sigma$ for simplification.

Given an information transmission policy $\Omega$, we define the corresponding information mapping $P_{\Omega}: L({\bf G})\rightarrow \Sigma^*$ recursively as follows:
\[
P_{\Omega}(\epsilon)=\epsilon; \\
  P_{\Omega}(s\sigma)  =
  \begin{cases}
  P_{\Omega}(s)\sigma, \mbox{ if}\ \Omega(s\sigma)=Y;\\
   P_{\Omega}(s), \mbox{ if}\ \Omega(s\sigma)=N.
   \end{cases}
\]
That is, $P_{\Omega}(s)$ is the observation of string $s$ under $\Omega$.


 In this work, instead of considering specific objectives, e.g., control or diagnosis, we consider a particular class of properties; it is called Information-State-based (IS-based) properties \cite{yin2016uniform} and defined as follows.

\begin{Definition}(IS-Based Property)
Given an automaton ${\bf G}$, an IS-based property $\varphi$ w.r.t. ${\bf G}$ is a function $\varphi:2^Q\to \{0,1\}$, where $\forall i\in 2^Q$, $\varphi(i)=1$ means that $i$ satisfies this property; $\varphi(i)=0$ otherwise.
We say that sublanguage $L\subseteq L({\bf G})$ satisfies $\varphi$ w.r.t. ${\bf G}$ and $\Omega$, which
is denoted by $L \models_{{\bf G}}^{\Omega} \varphi$, if $\forall s \in L : \varphi(R_G(s,L)) = 1$, where $R_G(s,L)=\{\delta(q_0,t)\in Q : \exists t\in L({\bf G})\text{ s.t. }P_{\Omega}(s)=P_{\Omega}(t)\} $.
\end{Definition} 

For more details about the notion of IS-based property, the reader is referred to \cite{yin2016uniform}, which shows that many important properties in the DES literature, e.g., distinguishability, diagnosability, opacity, detectability, and
safety, can be formulated as IS-based properties. 

Our objective is to synthesize an information transmission policy such that any given property holds. 
We define the Information Transmission Problem (IT Problem) as follows.
\begin{Porblem}\label{problem1}
 Given a plant $\textbf{G}=(Q,\Sigma,\delta,q_0)$  and an  IS-based property $\varphi:2^Q\to \{0,1\}$.  Find an information transmission policy $\Omega=({\bf A}, \mathcal{L})$ s.t.  $\varphi$ is satisfied  w.r.t. ${\bf G}$ and $\Omega$, i.e., 
 
 $$L({\bf A})\cap L({\bf G}) \models_{{\bf G}}^{\Omega} \varphi.$$
 
 \end{Porblem}

To this end, our approach proposed underlying the framework of DESs are given in the remainder of the paper: at first, we list all feasible information transmission policies (Section~\ref{sec.3}); then, according to specific rules, we extract a procedure that solves the IT Problem from these feasible policies (Section~\ref{sec.4}).

\section{A general most comprehensive dynamic observer}\label{sec.3}

In this section, for implementation purposes, we first attach the information transmission labels $\{Y, N\}$ to each state of the system ${\bf G}$. Then, a dynamic observer is constructed so as to contain all possible information transmission policies. We formally show that the information updating rule of the dynamic observer indeed yields the state estimate of the receiver.



\subsection{Labeled system}\label{sec.3.1}
We attach the information transmission labels to the system \textbf{G}, called \emph{labeled system}, which can be represented as a nondeterministic finite-state automaton (NFA)
\begin{align}\label{GAG}
{\bf G}_{ag}=(\tilde{Q},\Sigma,f,\tilde{Q}_{0}),
\end{align}
where 
\begin{itemize}
    \item 
    $\tilde{Q}=Q\times \{Y, N\}^{|\Sigma|}$ is the set of states;
    \item 
    $\Sigma$ is the  set of events;
    \item 
    $f:\tilde{Q}\times \Sigma  \to 2^{\tilde{Q}}$ is the partial transition function defined by: 
    for any $\tilde{q},\tilde{q}'\in \tilde{Q}$ and $\sigma\in \Sigma$, $\tilde{q}'\in f(\tilde{q},\sigma)$ iff 
    $q'=\delta(q,\sigma)$. The labels of $\tilde{q}$ are free to change, i.e., $\tilde{q}\in \{ q \} \times \{Y, N\}^{|\Sigma|}$.
    \item 
    $\tilde{Q}_0=\{ q_0 \} \times \{Y, N\}^{|\Sigma|}$. 
\end{itemize}
As shown above, the state space of ${\bf G}_{ag}$ is defined by $\tilde{Q}=Q\times \{Y, N\}^{|\Sigma|}$. For each state $q\in Q$, we list the possible transmission decisions of each event in $\Sigma$. Specifically, for an event $\sigma$, we have two transmission decisions: transmit event $\sigma$ (labeled by $Y$) and not transmit event $\sigma$ (labeled by $N$) at state $q$. We employ $\tilde{q}\in \{ q \} \times \{Y, N\}^{|\Sigma|}$ to show these different transmission decisions.   The transmission decision of $\tilde{q}$ for event $\sigma$ can be presented by $Lab_{\sigma}(\tilde{q})\in \{Y,N\}$.  The computational complexity of ${\bf G}_{ag}$ thus be $|Q|\cdot 2^{|\Sigma|}$. Hereafter in the paper, instead of listing all the labels of events in $\Sigma$, for $\tilde{q}$ we only list the labels of events that are defined at $q$, where the set of events defined at $q$ is given by $\Sigma_{q}=\{\sigma\in \Sigma|\delta(q,\sigma)!\}$. 
An illustrative example for the labeled system is given in the following.

\begin{figure}
  \centering
  \includegraphics[width=0.4\textwidth]{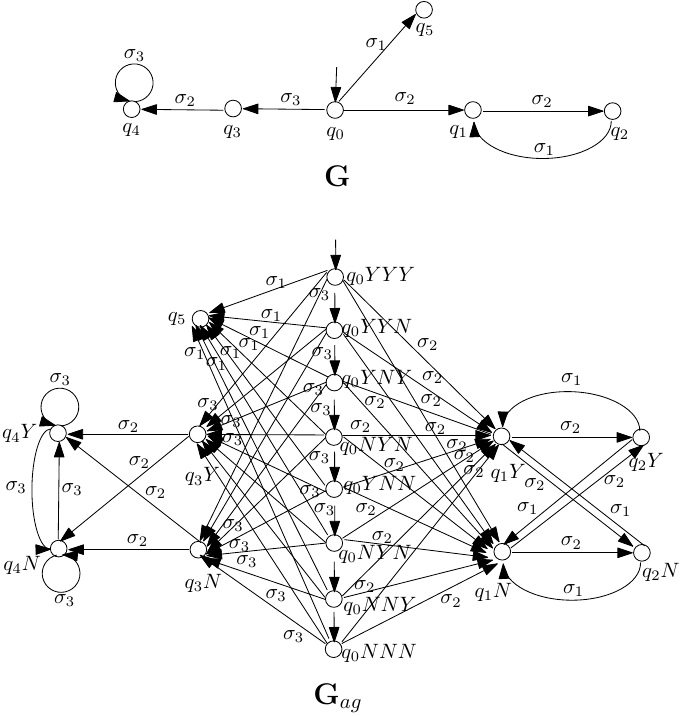}\\
  \caption{System ${\bf G}$ and labeled system ${\bf G}_{ag}$. }
  \label{fig:G1}
\end{figure} 


\begin{Example}An illustrative example is given in Fig.~\ref{fig:G1}. Given a plant ${\bf G}$, the construction ${\bf G}_{ag}$ can be built in Fig.~\ref{fig:G1} by attaching the information transmission labels $\{Y,N\}$ to states of ${\bf G}$, which is defined by (\ref{GAG}). For instance, the initial state of ${\bf G}_{ag}$ is $\tilde{Q}_0=\{ q_0 \} \times \{Y, N\}^{|\Sigma_{q_0}|}=\{ q_0 \} \times \{Y, N\}^3=\{q_0YYY,q_0YYN,q_0YNY,q_0NYY,\allowbreak q_0YNN,\allowbreak q_0NNY, \allowbreak q_0NNY,\allowbreak q_0NNN\}$ since there have three events ($\sigma_1$, $\sigma_2$, and $\sigma_3$) to be defined at state $q_0$. Accordingly, the state space of ${\bf G}_{ag}$ can be obtained by $\tilde{Q} =\allowbreak \{\allowbreak q_0YYY,\allowbreak q_0YYN,\allowbreak q_0YNY,\allowbreak q_0NYY,\allowbreak q_0YNN,\allowbreak q_0NNY,\allowbreak q_0NNY, \allowbreak q_0NNN, \allowbreak q_1Y,\allowbreak q_1N,\allowbreak q_2Y,\allowbreak q_2N,\allowbreak q_3Y,\allowbreak q_3N,\allowbreak q_4Y,\allowbreak q_4N,\allowbreak q_5\}.\allowbreak$ \hfill$\diamond$\end{Example}\label{ex1}

\subsection{Dynamic observer}\label{sec.3.2}

In order to capture all possible information transmission policies, a so-called dynamic observer is necessary to be constructed for ${\bf G}_{ag}$.
To proceed, some definitions are provided as follows.

The notation $\mathbb{L}$ is first employed to denote the set of all functions $l:\Sigma\rightarrow \{Y,N\}$ for the purpose of representing the transmission policy of each defined event, where the function $l$ maps events in $\Sigma$ to the labels $\{Y,N\}$. Specifically, $l(\sigma)=Y$ means the event $\sigma$ is transmitted by the sensor, and $l(\sigma)=N$ represents the opposite. 
In the usual way, $l$ is extended to $l:\Sigma^*\rightarrow \{Y,N\}^*$. 
A projection $\tilde{P}: L\rightarrow \Sigma^*$ is recursively defined by
\[
\tilde{P}(\epsilon)=\epsilon; \\
  \tilde{P}(s\sigma)  =
  \begin{cases}
  \tilde{P}(s)\sigma, \mbox{ if}\ l(s\sigma)=l(s)Y;\\
   \tilde{P}(s), \mbox{ if}\ l(s\sigma)=l(s)N.
   \end{cases}
\]

 The set of states that are unobservable reached from  $\tilde{q}$ thus can be obtained by 
\begin{align}\label{sq}
     S(\tilde{q})=\{\tilde{q}'\in \tilde{Q}|\forall s\in \Sigma^*\ s.t.\ \tilde{q}'\in f(\tilde{q},s)\wedge \tilde{P}(s)=\epsilon\} 
\end{align}
$S(\tilde{q})$ also can be written as $S(\tilde{q})=\bigcup_{(\forall s\in \Sigma^*)\ \tilde{P}(s)=\epsilon }f(\tilde{q},s)$.
We thus have $S(\imath)=\{ S(\tilde{q}):\tilde{q}\in\imath\}.$

A set of states $\imath\subseteq \tilde{Q}$ is said to be \emph{unobservable reach-closed} if for any states $\tilde{q}\in  \imath$ and any event $\sigma\in \Sigma^*$, we have that
\begin{align*}
\tilde{P}(\sigma)=\epsilon\ \wedge\ f(\tilde{q},\sigma)!\ \implies \exists\ \tilde{q}'\in \imath\ s.t.,\ \tilde{q}'\in f(\tilde{q},\sigma).
\end{align*}
That is to say, for any state $\tilde{q}$ in $\imath$, if there exists a nonempty string $\sigma$ that is defined at $\tilde{q}$ and not be transmitted, i.e., $\tilde{P}(\sigma)=\epsilon$, then there must exists a state $\tilde{q}'$ in $\imath$ such that $\tilde{q}'$ is reachable by $\sigma$ from $\tilde{q}$.

Consider the collection of all subsets of $ S(\tilde{q})$ that are unobservable reach-closed:
\begin{align*}
  \mathcal{UC}(\tilde{q})=\{\jmath\subseteq S(\tilde{q}):   \jmath \mbox{ is unobservable reach-closed}\}.
\end{align*}
It is straightforward to verify that $\mathcal{UC}(\tilde{q})$ is nonempty (\{$\tilde{q}\}$ belongs to) and is closed
under arbitrary unions. 

Let $\imath=\{\tilde{q}_1,\dots,\tilde{q}_n\}$. The subsets of $\imath$ that are unobservable reach-closed are defined by 
\[
\mathcal{UC}(\imath)=\bigcup_{\tilde{q}\in \imath}\mathcal{UC}(\tilde{q}).
\]
Moreover, there may be a decision conflict because for each $\sigma\in \Sigma$ we can choose either $Y$ or $N$. 
That is, some states cannot occur simultaneously in $ S(\tilde{q})$.
We say that a set $\imath\subseteq  S(\tilde{q})$ is non-conflicting if for any states $\tilde{q}_1, \tilde{q}_2\in  \imath$ and $i=1,2$,
 $\forall\ s_i\in \Sigma^*$ with $\tilde{q}_i\in f(\tilde{q}_0,s_i):\tilde{q}_1\neq \tilde{q}_2\ \Rightarrow\  s_1\neq~s_2$.

Similarly, consider the collection of all subsets of $ S(\tilde{q})$ that are non-conflicting:
\[
\mathcal{NC}(\tilde{q})=\{\jmath\subseteq  S(\tilde{q}): \jmath\ \mbox{is non-conflicting}\}.
\]
It can be verified that $\mathcal{NC}$ is nonempty and is closed under arbitrary intersections. 
Similarly, let $\imath=\{\tilde{q}_1,\dots,\tilde{q}_n\}$. The subsets of $\imath$ that are non-conflicting are defined by 
\[
\mathcal{NC}(\imath)=\bigcup_{\tilde{q}\in \imath}\mathcal{NC}(\tilde{q}).
\]
Let
\begin{align}\label{maxs0}
maxS(\imath)=\mathcal{UC}(\imath)\cap \mathcal{NC}(\imath)
\end{align}
be the maximal subset of $S(\imath)$ that is unobservable reach-closed and non-conflicting. Note that $maxS(\imath)$ is not a singleton because there may have different maximal choices that are conflicting. It can be verified that $maxS(\imath)=\mathcal{UC}(\mathcal{NC}(\imath))=\mathcal{NC}(\mathcal{UC}(\imath))$.


Let $\sigma\in \Sigma$ be an observable event. 
We have 
\begin{multline}\label{ux}
    N\!X_{\sigma} (\tilde{q})=\{\tilde{q}'\in \tilde{Q}: \exists \ s\in \Sigma^* \exists \tilde{q}_0\in \tilde{Q}_0\  s.t. \ f(\tilde{q}_0,s\sigma) \\ =f(\tilde{q},\sigma)\wedge
l(s\sigma)=l(s)Y \wedge \tilde{q}'\in f(\tilde{q},\sigma)\}
\end{multline}
For $\imath= \{\tilde{q}_1,\dots, \tilde{q}_n\}$, we define 
\begin{align}\label{next}
N\!X_{\sigma} (\imath)=\bigcup_{\tilde{q}\in \imath}N\!X_{\sigma} (\tilde{q}).
\end{align}


Now we are ready to construct the dynamic observer of ${\bf G}_{ag}$, which is defined as a new NFA
\begin{align}\label{observer g}
Obs({\bf G}_{ag})=(Z,\Sigma,\xi,Z_{0}),
\end{align}
where 
\begin{itemize}
    \item 
$Z\subseteq 2^{\tilde{Q}}$ is the set of states;

\item
$\Sigma$ is the set of events;

\item
$\xi: Z\times \Sigma \to 2^Z$ is the partial transition function defined by: 
for any $z\in Z, \sigma\in \Sigma$, we have 
\begin{flalign}\label{maxs}
& \xi(z,\sigma)= maxS( N\!X_{\sigma} (z ))=\bigcup_{z'\in N\!X_{\sigma} (z )}maxS(z'); &
\end{flalign}

\item
$Z_0=maxS(\tilde{Q}_0)$ is the set of initial states. 
 \end{itemize}

The initial states $Z_0$ is defined by a set of states from $\tilde{Q}_0$, i.e., $Z_0=maxS(\tilde{Q}_0)=\bigcup_{\tilde{q}\in\tilde{Q}_0}maxS(\tilde{q})$.
For state $z\in Z$, $maxS(z)$ is the set of states that can be reached unobservably from some state in $z$; it is also unobservable reach-closed and non-conflicting.  
Since $maxS(z)$ may not be a singleton, $N\!X_{\sigma} (z)$ is the set of states that can be reached from some state $\tilde{q}$ in $z$ immediately by a transmitted event $\sigma$, i.e., $l(\sigma)=Y$ and $f(\tilde{q},\sigma)!$. Similarly, $f(\tilde{q},\sigma)$ may not be a singleton since the future states of $\tilde{q}$ reached by $\sigma$ are attached with different labels.

As defined above, we employ $ N\!X_{\sigma}(z)$ to update the states when the event $\sigma$ is transmitted at state $z\in Z$. After obtaining the updated states, we use $maxS(N\!X_{\sigma} (z))$ to compute the states that are reached unobservably from states in $ N\!X_{\sigma} (z)$. 
Since $maxS(N\!X_{\sigma} (z))=\mathcal{UC}(N\!X_{\sigma} (z))\cap \mathcal{NC}(N\!X_{\sigma} (z))$, we first need to find all the states that are reached unobservably from states in $N\!X_{\sigma} (z)$. 
Then, by deleting the states that are not unobservable reach-closed and conflicting, we obtain the states that satisfy (\ref{maxs0}). 
It is ensured that all states in $Obs({\bf G}_{ag})$ are unobservable reach-closed and non-conflicting.
In addition, $Obs({\bf G}_{ag})$ contains all feasible information transmission policies.
According to the definition $Obs({\bf G}_{ag})$, the computational complexity of $Obs({\bf G}_{ag})$ is $2^{|Q|\cdot 2^{|\Sigma|}}$ with respect to the number of states $|Q|$ and the number of events $|\Sigma|$ in the system {\bf G}.
An illustrative example is given in the following. 

\begin{Example}
Let us  consider the system {\bf G} and the labeled system ${\bf G}_{ag}$ in Fig.  \ref{fig:G1}. We employ this example to illustrate the procedure of synthesizing the state set of $Obs({\bf G}_{ag})$. In this example, we only show partial paths started from one of the initial state $q_0NNY\in \tilde{Q}_0$ in the labeled system ${\bf G}_{ag}$. The cases started from other initial states are similar.

\begin{enumerate}
\item[(1)]
Initially, the initial states of $Obs({\bf G}_{ag})$ is computed from $q_0NNY$ in terms of that $maxS(q_0NNY) = \mathcal{UC}(q_0NNY) \cap \mathcal{NC}(q_0NNY)$.
More specifically, by (\ref{sq}), we have that $S(q_0NNY)\allowbreak =\{f(q_0NNY,\sigma_1),\allowbreak f(q_0NNY,\sigma_2),f(q_0NNY,\sigma_2\sigma_2),f(q_0NNY,\sigma_2(\sigma_2\allowbreak\sigma_1)^*)\}=\{q_0NNY, q_5, q_{1}Y, q_{1}N, q_{2}Y, q_{2}N\}$ for~the case that $ \tilde{P}(\sigma_1)\allowbreak=\allowbreak\tilde{P}(\sigma_2)\allowbreak=\allowbreak\tilde{P}(\sigma_2\sigma_2)\allowbreak=\allowbreak\epsilon$. 
Then, one can obtain that $\mathcal{UC}(q_0NNY)\cap \mathcal{NC}(q_0NNY)=\{\{q_0NNY,\allowbreak q_5,\allowbreak q_{1}Y\},\allowbreak \{q_0N\allowbreak NY,\allowbreak q_5,\allowbreak q_{1}N,\allowbreak q_{2}Y\},\{q_0N\allowbreak NY,q_5, q_{1}N,\allowbreak q_{2}N\},\allowbreak\{q_0NNY,\allowbreak q_5,\allowbreak q_{1}N,\allowbreak q_{2}N,\allowbreak q_{1}Y\},\allowbreak \{q_0NNY,\allowbreak q_5,\allowbreak q_{1}N,\allowbreak q_{2}N,\allowbreak q_{2}Y\}\}$.
Note that $\mathcal{UC}(q_0N\allowbreak NY)\allowbreak\cap \allowbreak\mathcal{NC}(q_0NNY)\allowbreak\subseteq \allowbreak 2^{S(q_0NNY)}$ since the rest subsets of $S(q_0NNY)$ are not unobservable reach-closed or/and non-conflicting. 
For example, the subset $\{q_0NNY,q_5,q_{1}N\}$ is not unobservable reach-closed  since there does not exist a state in this set such that $q_{1}N$ can reach it; the subset $\{q_0NNY,q_5,q_{1}N, q_{1}Y\}$ is conflicting because $q_{1}N$ and $q_{1}Y$ can be reached by the same event $\sigma_2$ from the initial state $q_0NNY$.
\item[(2)]
Then, the initial state $z_0=\{q_0NNY,\allowbreak q_5,\allowbreak q_{1}Y\}$ in $maxS(q_0NNY)$ is employed to illustrate how to update the states in  $Obs({\bf G}_{ag})$. 
Since there have two events $\sigma_2,\sigma_3$ that are defined at $\{q_0NNY,q_5,q_{1}Y\}$, the dynamic observer can evolve by (\ref{maxs}) as follows:
\item[(2-i)]
 $\xi(z_0,\sigma_2)= U\!R( N\!X_{\sigma_2} (z_0))=\bigcup_{z'\in N\!X_{\sigma_2} (z_0)}\allowbreak max\allowbreak S(z')$. 
 In this case, we have that $ N\!X_{\sigma_2} (z_0)=\{\{\tilde{q}_2'\}:\tilde{q}_2'\in f(q_0NNY,\sigma_2)\}\allowbreak =\allowbreak \{\{q_2Y\},\allowbreak \{q_2N\}\}$ according to (\ref{next}). Then, $\xi(z_0,\sigma_2)=maxS(\{q_2Y\})\cup maxS(\{q_2N\})$ is obtained, where $\{q_2Y\}$ is directly captured by $maxS(\{q_2Y\})\allowbreak=\{q_2Y\}$ since it is labeled with $Y$.
 Regarding to state $\{q_2N\}$, we can first obtain that $S(\{q_2N\})=\{f(q_2N,\sigma_1),f(q_2N,(\sigma_1\sigma_2)^*),\allowbreak f(q_2N,(\sigma_1\sigma_2)^*\allowbreak\sigma_2), f(q_2N,\sigma_1(\sigma_1\sigma_2)^*\sigma_2)\}=\{q_1N,\allowbreak \allowbreak q_{2}N,\allowbreak q_1Y,\allowbreak q_2Y\}$. 
 Then, we have that $maxS(\{q_2N\})=\{\{q_2N,q_1Y\},\{q_1N,q_{2}N\},\{q_1N,q_{2}N,q_1Y\}, \{q_1N,\\ q_{2}N,q_2Y\}\}\subseteq S(\{q_2N\}$ via (\ref{maxs}).
 Thus, we conclude with that $\xi(z_0,\sigma_2)=maxS(\{q_2Y\})\allowbreak \cup \allowbreak maxS(\{q_2N\})\allowbreak =\allowbreak \{\{q_2Y\},\allowbreak \{q_2N,\allowbreak q_1Y\},\allowbreak \{q_1N,\allowbreak q_{2}N\},\allowbreak \{q_1N,~q_{2}N,\allowbreak q_1Y\},  \{q_1N,\allowbreak q_{2}N,\allowbreak q_2Y\}\}$.
 
 \begin{figure}[t]
  \centering  \includegraphics[width=0.5\textwidth]{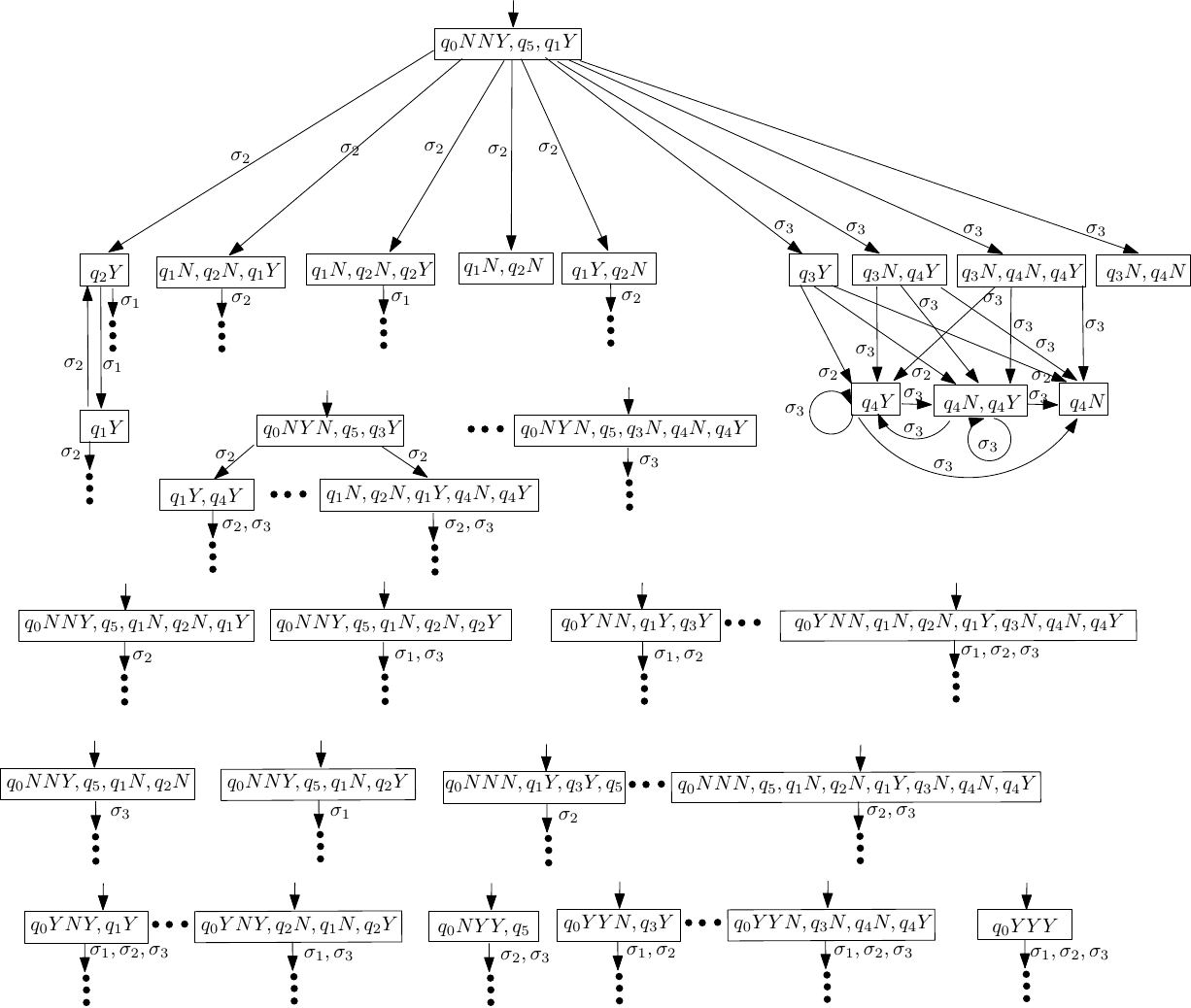}\\
  \caption{The partial dynamic observer $Obs({\bf G}_{ag})$.}
  \label{fig:G2}
\end{figure}

\item[(2-ii)] 
$\xi(z_0,\sigma_3)= U\!R( N\!X_{\sigma_3} (z_0))=\bigcup_{z'\in N\!X_{\sigma_3} (z_0)}\allowbreak max\allowbreak S(z')$. 
In this case, we have that $ N\!X_{\sigma_3} (z_0)=\{\{\tilde{q}_3'\}:\tilde{q}_3'\in f(q_0NNY,\sigma_3)\}=\{\{q_3Y\},\{q_3N\}\}$ according to (\ref{next}).
In a manner similar to the case (i), $S(\{q_3N\})=\{f(q_3N,\sigma_2),f(q_3N,\sigma_2\sigma_3^*)\}=\{q_3N,q_4N,q_4Y\}$ can be obtained. 
Then, we have that $maxS(\{q_3N\})=\{\{q_3N,\allowbreak q_4Y\},\allowbreak \{q_3N,\allowbreak q_4N\},\allowbreak \{q_3N,\allowbreak q_4N,\allowbreak q_4Y\}\}\allowbreak \subseteq \allowbreak S(\{q_3N\})$.
Thus, we conclude that $\xi(z_0,\allowbreak \sigma_3)\allowbreak =maxS(\{q_3Y\})\allowbreak \cup \allowbreak maxS(\{q_3N\})\allowbreak =\{\{q_3Y\},\allowbreak \{q_3N,\allowbreak q_4Y\},\allowbreak \{q_3N,\allowbreak q_4N\},\allowbreak \{q_3N,\allowbreak q_4N, q_4Y\}\}$.
\end{enumerate}

In the end, the partial dynamic observer $Obs({\bf G}_{ag})$ shown in Fig.~\ref{fig:G2} is obtained.
By following the above procedure, the rest states in $Obs({\bf G}_{ag})$ can be computed in a similar manner.
\hfill
$\diamond$
\end{Example}

\subsection{State Estimate}

Given a system ${\bf G}$ and an information transmission policy $\Omega=({\bf A}, \mathcal{L})$, in this subsection, we prove that the state estimate of ${\bf G}$ under  $\Omega$ is consistent with  the dynamic observer of $Obs({\bf G}_{ag})$.

For any string $s\in L({\bf G})$  generated by the system, we define
\begin{align}
 \mathcal{E}_{\Omega}^G(s):=\{\delta(q_0,t)\in Q : \exists t\in L({\bf G})\text{ s.t. }P_{\Omega}(s)=P_{\Omega}(t)\} \label{est}
\end{align}
as the {\it state  estimate} of the receiver. 
Clearly, for  strings $s,t\in L({\bf G})$, if $P_{\Omega}(s)=P_{\Omega}(t)$, then   $\mathcal{E}_{\Omega}^G(s)=\mathcal{E}_{\Omega}^G(t)$. 
The reverse information mapping is defined as $P_{\Omega}^{-1}(s)=\{s'\in \Sigma^*: P_{\Omega}(s)=P_{\Omega}(s')\}$.

In order to restrict the information transmission policies to a specific policy $\Omega=({\bf A}, \mathcal{L})$,  we construct the product of ${\bf A}=(X,\Sigma,\eta,x_{0})$ and ${\bf G}_{ag}=(\tilde{Q},\Sigma,f,\tilde{Q}_{0})$ by ${\bf A}\times {\bf G}_{ag}=(V,\Sigma,\theta,v_0)$, where 
\begin{itemize}
    \item 
$V\subseteq X\times\tilde{Q}$ is the set of states, 

\item
$\Sigma$ is the set of events, 

\item
$\theta: V\times \Sigma \to V$ is the partial transition function defined by: 
for any $v=(x,\tilde{q}), v'=(x',\tilde{q}')\in V$, and $\sigma\in \Sigma$, $\forall s\sigma'\in \Sigma^*$ with $f(\tilde{q}_0,s\sigma')=f(\tilde{q}',\sigma')$ and $\eta (x',\sigma')!$, we have 
that $v'=\theta(v,\sigma)$ iff
\begin{align}\label{eta}
x'=\eta (x,\sigma)\ \wedge\ \tilde{q}'\in f(\tilde{q},\sigma)\ \wedge\ l(s\sigma')=l(s)\mathcal{L}(x',\sigma');
\end{align}

\item
$v_0=(x_0,\tilde{q}_0)$ with $l(\sigma)=\mathcal{L}(x_0,\sigma)$ is the initial state when $f(\tilde{q}_0,\sigma)!$ and $\eta(x_0,\sigma)!$ for all $\sigma\in \Sigma$.
 \end{itemize}
 
To estimate the states of the system under the given information transmission policy $\Omega$, we construct the observer of ${\bf A}\times {\bf G}_{ag}$, which is defined by $Obs({\bf A}\times {\bf G}_{ag})=(H,\Sigma,\gamma,h_0)$, where 
 \begin{itemize}
    \item 
$H\subseteq 2^V=2^{X\times \tilde{Q}}$ is the set of states, 

\item
$\Sigma$ is the set of events, 

\item
$\gamma: H\times \Sigma \to H$ is the partial transition function defined by: 
for any $h, h'\in H, \sigma\in \Sigma$, we have 
$h'=\gamma(h,\sigma)$ iff
\begin{equation}\label{nur}
h'=UR'(NX'_{\sigma}(h))
\end{equation}
where for any $h\in 2^V$, we have
\begin{equation}\label{nextn}
N\!X'_{\sigma} (h)=\{v'\in V:\exists\ v\in h\ s.t.\ \theta(v,\sigma)=v'\}
\end{equation}
\begin{align*}
& U\!R'(h)=\bigcup_{v\in h}U\!R'(v)
\end{align*}
\begin{multline}\label{ur}
 U\!R'(v)=\{v'\in V: \ \exists\ s\in \Sigma^*\ s.t.\\P_{\Omega}(s)=\epsilon\ \wedge \ v'=\theta(v,s)\},
\end{multline}
\item
$h_0=UR'(v_0)$ is the initial states. 
 \end{itemize}
Next, we show that the labeled state components of the observer $Obs({\bf A}\times {\bf G}_{ag})$ under an arbitrary policy $\Omega$ always belongs to the state set of the dynamic observer $Obs({\bf G}_{ag})$ after an observable string, which shows that our synthesized dynamic observer $Obs({\bf G}_{ag})$ contains all possible state estimation for the system.  
For any $h\in H$, let
\[
I_2(h)=\{\{\tilde{q}_1,\dots,\tilde{q}_m\}\in 2^{\tilde{Q}}: h=(x_1,\dots,x_k,\tilde{q}_1,\dots,\tilde{q}_m)\}.
\]

\begin{Proposition}\label{pro.1}
Let $Obs({\bf G}_{ag})=(Z,\Sigma,\xi,Z_{0})$ be the dynamic observer  induced  by (\ref{observer g}), and $s=\sigma_1\sigma_2\dots \sigma_n$ be an observable string available to the observer $Obs({\bf A}\times {\bf G}_{ag})=(H,\Sigma,\gamma,h_0)$. Then, we have 
\begin{align*}
I_2(\gamma(h_0,s))\subseteq\bigcup_{z_0\in Z_0}\xi(z_0,s).
\end{align*}
\end{Proposition}
\textbf{Proof.} Assume that there exists a state $\imath\in H$ s.t. $I_2(\imath)=I_2(\gamma(h_0,s))$ and $\imath=\gamma(h_0,s)$.
We have the following:
\begin{align*}
& \imath=\gamma(h_0,s)\\
&=\gamma(h_0,\sigma_1\sigma_2\dots \sigma_n)\\
&=\gamma(\dots\gamma(\gamma(h_0,\sigma_1),\sigma_2)\dots ),\sigma_n)\\
&=\gamma(\dots\gamma(UR'(NX'_{\sigma_1}(h_0)),\sigma_2)\dots ),\sigma_n) \mbox{\ by\ (\ref{nur})}\\
&=\gamma(\dots\gamma(UR'(NX'_{\sigma_1}(UR'(v_0))),\sigma_2)\dots ),\sigma_n)\\
&=\gamma(\dots\gamma(UR'(NX'_{\sigma_1}(\bigcup_{\substack{\forall s_0\in \Sigma^*| \\ P_{\Omega}(s_0)=\epsilon}}\theta(v_0,s_0))),\sigma_2)\dots ),\sigma_n)\\
&\mbox{ by\ (\ref{ur})}\\
&=UR'(NX'_{\sigma_n}(\dots (UR'(\bigcup_{\substack{\forall s_0\in \Sigma^*|\\ P_{\Omega}(s_0)=\epsilon}}\theta(v_0,s_0\sigma_1)))\dots ))\\
& \mbox{with\ $l(s_0\sigma_1)=\mathcal{L}(x_0,s_0\sigma_1)$ by\ (\ref{eta}) }\\
& =UR'(NX'_{\sigma_n}(\dots (\bigcup_{\substack{\forall s_1\in \Sigma^*|\\ P_{\Omega}(s_1)=\epsilon}}\bigcup_{\substack{\forall s_0\in \Sigma^*|\\ P_{\Omega}(s_0)=\epsilon}}\theta(v_0,s_0\sigma_1s_1))\dots ))  \\
&=\bigcup_{\substack{\forall s_n\in \Sigma^*|\\ P_{\Omega}(s_n)=\epsilon}}\dots \bigcup_{\substack{\forall s_1\in \Sigma^*|\\ P_{\Omega}(s_1)=\epsilon}}\bigcup_{\substack{\forall s_0\in \Sigma^*|\\ P_{\Omega}(s_0)=\epsilon}}\theta(v_0,s_0\sigma_1s_1\dots s_{n-1}\sigma_n s_{n})\\
& \mbox{with\ $l(s_0\sigma_1s_1\dots s_{n-1}\sigma_n s_n)=\mathcal{L}(x_0,s_0\sigma_1s_1\dots s_{n-1}\sigma_n s_n)$} \\
&=\bigcup_{\substack{\forall s_n\in \Sigma^*|\\ P_{\Omega}(s_n)=\epsilon}}\dots \bigcup_{\substack{\forall s_0\in \Sigma^*|\\ P_{\Omega}(s_0)=\epsilon}}\theta((x_0,\tilde{q}_0),s_0\sigma_1s_1\dots s_{n-1}\sigma_n s_{n}).\\
& \mbox{Then, by\ (\ref{eta}), we have that}\\
& I_2(\imath)=\bigcup_{\substack{\forall s_n\in \Sigma^*|\\ P_{\Omega}(s_n)=\epsilon}}\dots \bigcup_{\substack{\forall s_0\in \Sigma^*|\\ P_{\Omega}(s_0)=\epsilon}}f(\tilde{q}_0,s_0\sigma_1s_1\dots s_{n-1}\sigma_n s_n)\\
& \mbox{and\ $l(s_0\sigma_1s_1\dots s_{n-1}\sigma_n s_{n})=\mathcal{L}(x_0,s_0\sigma_1s_1\dots s_{n-1}\sigma_n s_{n})$.}
\end{align*}
Due to the fact that $Obs({\bf A}\times {\bf G}_{ag})$ is a DFA, one can obtain that $I_2(\imath)$ is non-conflicting since different state in DFA are reached by different strings; this is not always satisfied in NFA.
By\ (\ref{nur}), we obtain that $I_2(\imath)$ contains all unobservable reach states, which implies that $I_2(\imath)$ is unobservable reach-closed.
\begin{align*}
& \bigcup_{z_0\in Z_0}\xi(z_0,s)=\bigcup_{z_0\in Z_0}\xi(z_0,\sigma_1\sigma_2\dots \sigma_n)\\
& =\bigcup_{z_0\in Z_0}\xi(\dots\xi(\xi(z_0,\sigma_1),\sigma_2)\dots, \sigma_n)\\
& =\bigcup_{z_0\in Z_0}maxS(NX_{\sigma_n}(\dots(maxS(NX_{\sigma_1}(z_0))\dots) \mbox{\ by\ (\ref{maxs})}\\
& =\bigcup_{z_0\in Z_0} maxS(NX_{\sigma_n}(\dots(maxS(\bigcup_{\tilde{q}\in z_0}NX_{\sigma_1}(\tilde{q})))\dots))\\ 
&=\bigcup_{z_0\in Z_0} maxS(NX_{\sigma_n}(\dots (maxS(\bigcup_{\tilde{q}\in z_0}f(\tilde{q},\sigma_1))\dots)) \mbox{\ by\ (\ref{ux})}\\
 & \mbox{\ Let\ }R_1=maxS(\bigcup_{\tilde{q}\in z_0}f(\tilde{q},\sigma_1)).\\
\end{align*}
 \begin{align*}
& R_1=\{\jmath_1\subseteq S(\bigcup_{\tilde{q}\in z_0}f(\tilde{q},\sigma_1)):\jmath_1 \mbox{ is unobservable reach-closed} \\
& \mbox{ and non-conflicting}\}\\
&=\{\jmath_1\subseteq \bigcup_{\substack{\forall t_1\in \Sigma^*|\\ \tilde{P}(t_1)=\epsilon}}\bigcup_{\tilde{q}\in z_0}f(\tilde{q},\sigma_1t_1):\jmath_1 \mbox{ is unobservable reach-}\\ 
& \mbox{closed and non-conflicting}\} \mbox{ by (\ref{maxs})}\\
& maxS(NX_{\sigma_2}(R_1))=\{\jmath_2\subseteq \bigcup_{\substack{\forall t_2\in \Sigma^*|\\ \tilde{P}(t_2)=\epsilon}}\bigcup_{\substack{\forall t_1\in \Sigma^*|\\\tilde{P}(t_1)=\epsilon}}\bigcup_{\tilde{q}\in z_0}f(\tilde{q},\sigma_1t_1\sigma_2t_2):\\
& \jmath_2 \mbox{ is unobservable reach-closed and non-conflicting}\}\\
& \vdots\\
& \bigcup_{z_0\in Z_0}\xi(z_0,s)\\
& =\bigcup_{z_0\in Z_0}\{\jmath_n\subseteq \bigcup_{\substack{\forall t_n\in \Sigma^*|\\ \tilde{P}(t_n)=\epsilon}}\dots\bigcup_{\substack{\forall t_1\in \Sigma^*|\\ \tilde{P}(t_1)=\epsilon}}\bigcup_{\tilde{q}\in z_0}f(\tilde{q},\sigma_1t_1\sigma_2\dots\sigma_nt_n):\\
& \jmath_n \mbox{ is unobservable reach-closed and non-conflicting}\}.\\
& =\bigcup_{\tilde{q}'_0\in \tilde{Q}_0}\{\jmath_n\subseteq \bigcup_{\substack{\forall t_n\in \Sigma^*|\\ \tilde{P}(t_n)=\epsilon}}\dots\bigcup_{\substack{\forall t_1\in \Sigma^*|\\ \tilde{P}(t_1)=\epsilon}}\bigcup_{\tilde{q}\in maxS(\tilde{q}'_0)}f(\tilde{q},\sigma_1t_1\sigma_2\dots\\
& \sigma_nt_n):\jmath_n \mbox{ is unobservable reach-closed and non-conflicting}\}.\\
& =\bigcup_{\tilde{q}'_0\in \tilde{Q}_0}\{\jmath_n\subseteq \bigcup_{\substack{\forall t_n\in \Sigma^*|\\ \tilde{P}(t_n)=\epsilon}}\dots\bigcup_{\substack{\forall t_1\in \Sigma^*|\\ \tilde{P}(t_1)=\epsilon}}\bigcup_{\substack{\forall t_0\in \Sigma^*|\\ \tilde{P}(t_0)=\epsilon}}f(\tilde{q},t_0\sigma_1t_1\sigma_2\dots\\
& \sigma_nt_n):\jmath_n \mbox{ is unobservable reach-closed and non-conflicting}\}.\\
\end{align*}
For any string $\ell \in\Sigma^*$, we have that $P_{\Omega}(\ell)=\tilde{P}(\ell)$ via $P_{\Omega}(\ell)=\epsilon$ and $\tilde{P}(\ell)=\epsilon$, which means that both $P_{\Omega}(\ell)$ and $\tilde{P}(\ell)$ transmit string $\ell$. 
 We thus obtain that $\bigcup_{z_0\in Z_0}\xi(z_0,s)=\bigcup_{\tilde{q}'_0\in \tilde{Q}_0}\{\jmath_n\subseteq \bigcup_{\forall t_n\in \Sigma^*|\tilde{P}(t_n)=\epsilon}\dots\allowbreak\bigcup_{\forall t_1\in \Sigma^*|\tilde{P}(t_1)=\epsilon}\bigcup_{\forall t_0\in \Sigma^*|\tilde{P}(t_0)=\epsilon}f(\tilde{q},t_0\sigma_1t_1\sigma_2\dots\sigma_nt_n): \jmath_n $ is unobservable reach-closed and non-conflicting$\}$, which
 directly implies that $I_2(\gamma(h_0,s))\subseteq\bigcup_{z_0\in Z_0}\xi(z_0,s).$
\hfill
$\diamond$

Finally, we formally show that the state space of the dynamic observer $Obs({\bf A}\times {\bf G}_{ag})$ indeed yields the state estimate of the receiver. For $\imath\in 2^{\tilde{Q}}$, let 
\begin{align}\label{i1}
    I_1(\imath)=\{q\in Q: \tilde{q}\in \imath\}.
\end{align}

\begin{Theorem}\label{main}
Let $\Omega=({\bf A},\mathcal{L})$ be an information transmission policy imposed on ${\bf G}$ and $s=\sigma_1\sigma_2\dots\sigma_m$ be an observable string available to the dynamic observer $Obs({\bf A}\times {\bf G}_{ag})=(H,\Sigma,\gamma,h_0)$. Then, we have that
\[I_1(I_2(\gamma(h_0,s)))=\mathcal{E}_{\Omega}^G(s). \]
\end{Theorem}

\textbf{Proof.}
By Proposition~1, we have that $I_2(\gamma(h_0,s))=\bigcup_{P_{\Omega}(s_n)=\epsilon}\dots \bigcup_{P_{\Omega}(s_0)=\epsilon}f(\tilde{q}_0,s_0\sigma_1s_1\dots s_{n-1}\sigma_n s_n)$. 
Then, we obtain that 
\begin{align*}
 &I_1(I_2(\gamma(h_0,s)))=\{q\in Q: \tilde{q}\in \bigcup_{\substack{\forall s_n\in \Sigma^*|P_{\Omega}=\epsilon}}\dots \\
&\bigcup_{\substack{\forall s_0\in \Sigma^*|P_{\Omega}=\epsilon}}f(\tilde{q}_0,s_0\sigma_1s_1\dots s_{n-1}\sigma_n s_n)\}  \mbox{ by (\ref{i1})}\\
& =\{q\in Q: (\forall s_i\in\Sigma^*,\ i\in[1,n]) \tilde{q}\in f(\tilde{q}_0,s_0\sigma_1s_1\dots \\
&s_{n-1}\sigma_n s_n) \wedge P_{\Omega}(s_i)=\epsilon\}\\
&=\{q\in Q: (\forall s_i\in\Sigma^*,\ i\in[1,n])\ \tilde{q}\in f(\tilde{q}_0,s_0\sigma_1s_1\dots\\
& s_{n-1}\sigma_n s_n) \wedge P_{\Omega}(s_0\sigma_1s_1\dots s_{n-1}\sigma_n s_n)=P_{\Omega}(s)\}\\
&=\{q\in Q: (\forall s_i\in\Sigma^*,\ i\in[1,n])\ q\in \delta(q_0,s_0\sigma_1s_1\dots \\
&s_{n-1}\sigma_n s_n) \wedge P_{\Omega}(s_0\sigma_1s_1\dots s_{n-1}\sigma_n s_n)=P_{\Omega}(s)\} \mbox{ by (\ref{GAG})}\\
& =\mathcal{E}_{\Omega}^G(s) \mbox{ by (\ref{est})}. 
\end{align*}
\hfill
$\diamond$

\begin{figure}[htbp]
  \centering
  \includegraphics[width=0.4\textwidth]{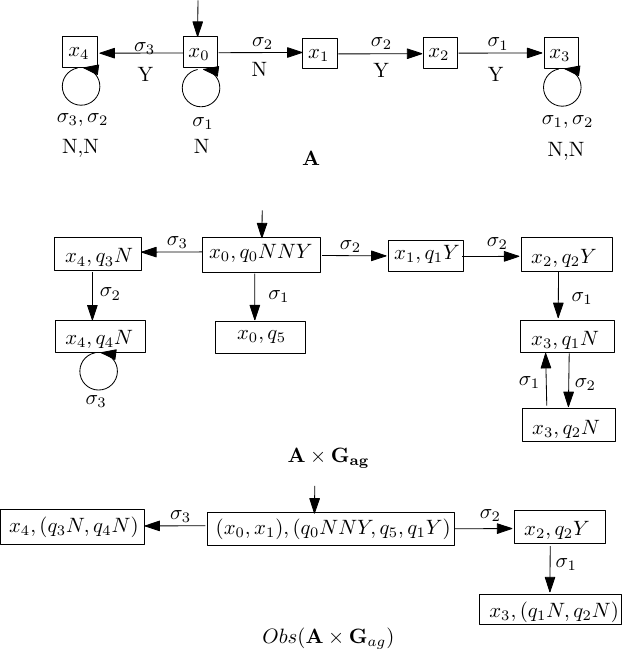}\\
  \caption{Sensor automaton ${\bf A}$, the product automaton ${\bf A}\times {\bf G}_{ag}$ and the observer $Obs({\bf A}\times {\bf G}_{ag})$. }
  \label{fig:G4}
\end{figure}

By Theorem~\ref{main}, we show that the state estimate of ${\bf G}$ under  $\Omega$ is consistent with  the dynamic observer of $Obs({\bf G}_{ag})$. 
\begin{Example}
Let us consider system ${\bf G}_{ag}$ in Fig.~\ref{fig:G1}. Given an information transmission policy $\Omega=({\bf A},\mathcal{L})$, where ${\bf A}$ is shown in Fig.~\ref{fig:G4} and $\mathcal{L}$ is given by: $\mathcal{L}(x_0,\sigma_1)=\mathcal{L}(x_0,\sigma_2)=\mathcal{L}(x_3,\sigma_1)=\mathcal{L}(x_3,\sigma_2)=\mathcal{L}(x_4,\sigma_3)=\mathcal{L}(x_4,\sigma_2)=N$ and $\mathcal{L}(x_0,\sigma_3)=\mathcal{L}(x_1,\sigma_2)=\mathcal{L}(x_2,\sigma_1)=Y$. The corresponding product automaton ${\bf A}\times {\bf G}_{ag}$ and the dynamic observer $Obs({\bf A}\times {\bf G}_{ag})$ defined above are given in Fig.~\ref{fig:G4}. Let string $s$  be an arbitrary observable string available to the  the dynamic observer $Obs({\bf A}\times {\bf G}_{ag})$. It can be verified that $I_2(\theta'(y_0',s))\subseteq\bigcup_{z_0\in Z_0}\xi(z_0,s)$ (Proposition~\ref{pro.1}) and $I_1(I_2(\theta'(y_0',s)))=\mathcal{E}_{\Omega}^G(s)$ (Theorem~\ref{main}) always hold.
\end{Example}

\section{Synthesis of the feasible transmission policy}\label{sec.4}
In this section, we discuss how to synthesize a deterministic information policy to ensure the IS-based property. Given the observer  $Obs({\bf G}_{ag})=(Z,\Sigma,\xi,Z_{0})$, we say that a state $z$ is \emph{consistent} if  $\forall \sigma\in \Sigma_z \neq \emptyset$, $NX_{\sigma}(z)\neq \emptyset$; it is \emph{inconsistent} otherwise. We denote by $Z_{const}$ the set of consistent states in $Z$ and we say that $Z$ is consistent if all reachable states are consistent.

Our approach for synthesizing a deterministic information policy, denoted by a DFA, consists of the following three steps:

Step (i) construct the largest sub-automaton ${\bf G}^*$ of $Obs({\bf G}_{ag})$ such that $L({\bf G}^*)$ satisfies the IS-based property and the states of ${\bf G}^*$ are consistent;

Step (ii) synthesize an automaton $\bf{A}=(X,\Sigma_A,\eta,x_{0})$ from the largest feasible automaton ${\bf G}^*$ to present one feasible information transmission policy, while ensures that the obtained automaton $\bf{S}$ transmits as fewer events as possible;

Step (iii) extract one deterministic information transmission policy $\Omega$ based on ${\bf S}^*$ to show how the events of ${\bf G}$ are transmitted at each step.
\subsection{Synthesis of the largest feasible automaton}\label{max_policy}
To satisfy the IS-based property $\varphi$, it should be guaranteed that, for any $\imath\in 2^{\tilde{Q}}$,
$\varphi(\imath)=1.$

To this end, we define
\[
Z_{dis}=\{\imath\in Z: \varphi(\imath)=0\}
\]
as the set of states that dissatisfies the IS-based property $\varphi$.

In order to synthesize a desired transmission information policy, we first construct the largest sub-automaton of $Obs({\bf G}_{ag})=(Z,\Sigma,\xi,Z_{0})$ that enumerates all the feasible transitions satisfying the constraints of $\xi$. Such an all-feasible automaton is denoted by  ${\bf G}_{total}$. Then,  we need to delete some states that violate the IS-based property and obtain a new automaton 

\[
{\bf G}_{0}={\bf G}_{total}\upharpoonright_{Z\setminus Z_{dis}},
\]
where ${\bf G}\upharpoonright_{Z}$ denotes the automaton obtained by restricting the state-space of ${\bf G}$ to $Z$.

After removing the states that violate the IS-based property, the resulting automaton may exist some inconsistent states, which should be removed in a recursive manner. 
To this end, we give an operator $R$ that maps an automaton to a new one by:
\[
R:{\bf G} \mapsto {\bf G}\upharpoonright_{Z_{const}}
\]
and define
\[
{\bf G}^*=\lim_{k\rightarrow\infty }R^k({\bf G}_{0})
\]
as the largest consistent automaton which satisfies the IS-based property. 

By following Algorithm~1, one is able to synthesize ${\bf G}^*$  via a depth-first search.
We briefly explain how Algorithm~1 works next.
By line 2, we first delete the states that violate the IS-based property. Then, by lines 4-8, we delete the inconsistent states. 
Finally, we obtain an automaton that satisfies the IS-based property by deleting all inconsistent states recursively. 
So far, the condition (i) of IT Problem in Section~II is satisfied.
\renewcommand{\algorithmicrequire}{\textbf{Input:}}
\renewcommand{\algorithmicensure}{\textbf{Output:}}
\begin{algorithm}[h]\label{alg.2}
\caption{Synthesis of the automaton ${\bf G}^*$.}
 \begin{algorithmic}[1]\Require$Obs({\bf G}_{ag})=(Z,\Sigma,\xi,Z_{0})$\Ensure${\bf G}^*=(\tilde{Z},\tilde{\Sigma},\tilde{\xi},\tilde{Z_{0})}$
\State   ${\bf G}_{total}=Obs({\bf G}_{ag})$;
\State   ${\bf G}^*={\bf G}_0 ={\bf G}_{total}\upharpoonright_{Z\setminus Z_{dis}}$;
\State  $Z_{const}=\tilde{Z}$;
\For{each $z\in \tilde{Z}$}
\If {$\exists \sigma\in \Sigma$ s.t. $\xi(z,\sigma)!$ and $\urcorner \tilde{\xi}(z,\sigma)!$ }
\State  $Z_{const}=Z_{const}\setminus \{z\}$;
\EndIf
\EndFor
\State      $R:{\bf G}_0 \mapsto {\bf G}_0\upharpoonright_{Z_{const}}$;
\State      ${\bf G}^*=\lim_{k\rightarrow\infty }R^k({\bf G}_{0})$;
    \end{algorithmic}
\end{algorithm}

\begin{figure}
  \centering
  \includegraphics[width=0.5\textwidth]{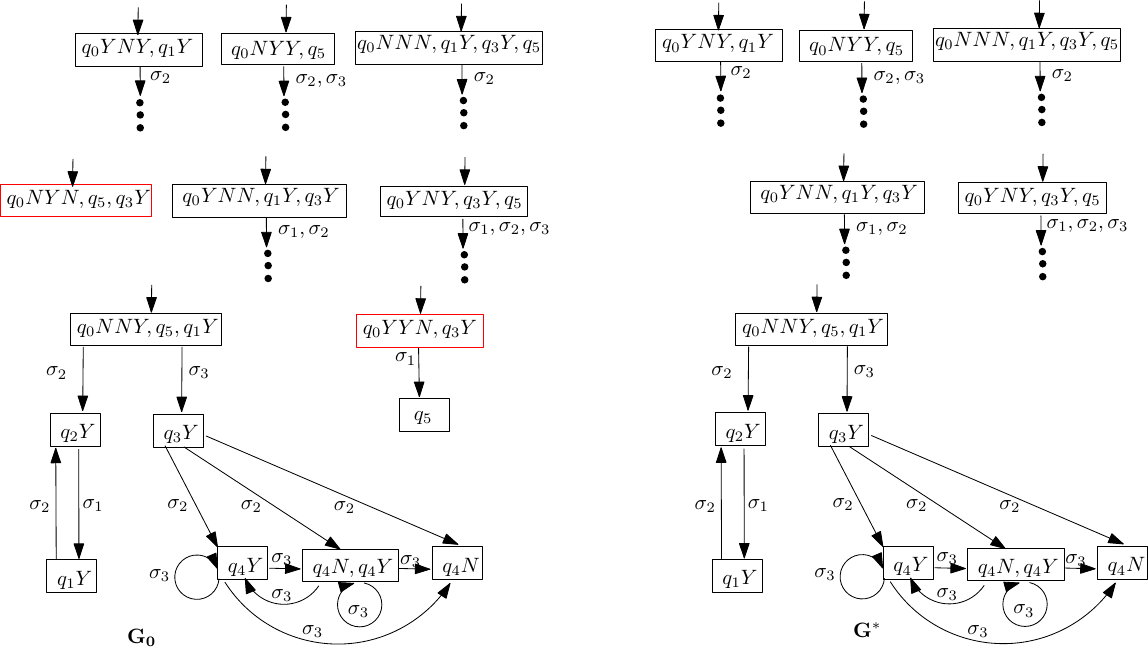}\\
  \caption{${\bf G}_0$ and ${\bf G}^*$ }
  \label{fig:G3}
\end{figure}

We employ one specific property, called \emph{distinguishability property}, to illustrate this result.
Let  $T\subseteq  Q\times Q$ be the specification imposed on the system  ${\bf G}=(Q,\Sigma, \delta, q_0)$, 
and $\Omega$ be the information transition policy. 
We say that ${\bf G}$ is distinguishable w.r.t. $\Omega$ and $T$ if for any string $s\in L({\bf G})$, we have that $(\mathcal{E}_{\Omega}^G(s)\times \mathcal{E}_{\Omega}^G(s)) \cap T =\emptyset.$
Note that $T$ can be expressed in terms of IS-based properties $\psi:2^Q\to \{0,1\}$ by 
\[
\forall i\in 2^Q:[ \psi (i)=0] \Leftrightarrow [\exists q_1,q_2\in i: (q_1,q_2)\in T].
\]
\begin{Example}
Let us  consider $Obs({\bf G}_{ag})$ in Fig.~\ref{fig:G1} and $T=  \{q_0,q_1,q_3,q_5\}\times \{q_2,q_4\}$. We first delete states that make the system ${\bf G}$ undistinguishable and obtain a distinguishable automaton ${\bf G}_0$ as shown in Fig.~\ref{fig:G3}. We note that there exist some states that are inconsistent in ${\bf G}_0$, e.g., $(q_0NYN,q_3Y,q_5)$ and $(q_0YYN,q_3Y)$. Event $\sigma_2$ is defined at states $(q_0NYN,q_3Y,q_5)$ and $(q_0YYN,q_3Y)$ in $Obs({\bf G}_{ag})$ and not defined in ${\bf G}_0$.  We thus need to delete inconsistent states recursively by Algorithm~\ref{alg.2} and obtain the largest consistent and distinguishable automaton ${\bf G}^*$ which is shown in Fig.~\ref{fig:G3}. 
\end{Example}

\subsection{Transmission policy with fewer transmitted events}
So far, the first step (Step (i)) has been finished. 
 Now, we execute Step (ii) by synthesizing an automaton $\bf{A}=(X,\Sigma_A,\eta,x_{0})$ from the largest feasible automaton ${\bf G}^*$ to present the corresponding information transmission policy, while ensures that the obtained automaton $\bf{A}$ transmits as fewer events as possible.

Since ${\bf G}^*$ is constructed by some sub-automaton with  different initial state, ${\bf G}^*=(\tilde{Z},\tilde{\Sigma},\tilde{\xi},\tilde{Z}_0)$ can be written as ${\bf G}^*=\{{\bf G}^*_1,\dots,{\bf G}^*_l\}$, where $l\leq 2^{|\Sigma_{q_0}|}$ and ${\bf G}^*_i=(\tilde{Z}_i,\tilde{\Sigma},\tilde{\xi}_i,\tilde{z}_{i0})$ is defined by
\begin{itemize}
    \item 
$\tilde{Z}_i\subseteq \tilde{Z}$ is the set of states and satisfies $\tilde{Z}=\tilde{Z}_1\cup\tilde{Z}_2\cup\dots\cup\tilde{Z}_l$;

\item
$\Sigma$ is the set of events;

\item
$\xi: \tilde{Z}_i\times \Sigma \to 2^{\tilde{Z}_i}$ is the partial transition function defined by: 
for any $z_i,z_j\in \tilde{Z}_i, \sigma\in \Sigma$, we have $z_j\in\tilde{\xi}_i(z_i,\sigma)$ iff
\begin{align}
   z_j\in\tilde{\xi}(z_i,\sigma);
\end{align}

\item
$z_{i0}\in \tilde{Z}_0$ is the set of initial states. 
 \end{itemize}
${\bf G}^*_i$ is the sub-automaton of ${\bf G}^*$ and has an unique initial state $z_{i0}$. The states of ${\bf G}^*_i$ can be reached from $z_{i0}$. 

To synthesize an automaton $\bf{S}=(Z',\Sigma,\xi',z_{0})$ from the largest feasible automaton ${\bf G}^*$ and ensure that ${\bf S}$ transmits as fewer events as possible, Algorithm 2 is proposed in the following.

\renewcommand{\algorithmicrequire}{\textbf{Input:}}
\renewcommand{\algorithmicensure}{\textbf{Output:}}
\begin{algorithm}[h]\label{alg.3}
\caption{Synthesis of the automaton ${\bf S}$.}
 \begin{algorithmic}[1]\Require${\bf G}^*=(\tilde{Z},\Sigma,\tilde{\xi},\tilde{Z_{0})}$, ${\bf G}^*_i=(\tilde{Z}_i,\tilde{\Sigma},\tilde{\xi}_i,\tilde{z}_{i0})$ for $i\in[1,l]$, $\tilde{Z}_i=\{z_{i0},z_{i1},\dots,z_{im}\}$
   \Ensure ${\bf S}=(Z',\Sigma,\xi',z_{0}')$
\For{$i=1$ to $i=l$}
 \State  $N_i=0$;
\For{each $z\in \tilde{Z}$}
   \State  $N_z=0$;
 \For{each $\tilde{q}\in z$}
 \If{$\exists\sigma\in\Sigma$ s.t. $\delta(\tilde{q},\sigma)\in z$}
      \State     $N_z=N_z+1;$
             \EndIf
     \EndFor
    \EndFor
 \State  $N_i=\Sigma_{z\in \tilde{Z}_i}N_z$;  
\EndFor
 \State  $k\in \{i:\forall j\in[1,l]\ N_i \geqslant N_j \}$;  
 \State  $z_0'=z_{k0}$;
\For{ $h=0$ to $h=m$}
\For{each $\sigma\in \Sigma$ with $\tilde{\xi}_k(z_{kh},\sigma)!$}
  \State $z_p\in\{z\in \tilde{Z}_i:\ (\forall z'\in NX'_{\sigma}(z_{kh}))\ N_z\geqslant N_{z'}\}$; 
  \State $\xi'(z_{kh},\sigma)=z_p$;
  \State $Z'=Z'\bigcup\{z_p\}$.
     \EndFor
   \EndFor
 \end{algorithmic} 
\end{algorithm}

 Algorithm~2 first computes the number of events that are not transmitted in each sub-automaton (lines 1-12). Then we choose the sub-automaton ${\bf G}^*_k$ which has the maximal number of non-transmitted events (lines 13-14). Note that there may exist more than one sub-automaton that has the maximal number of non-transmitted events, we choose one of them by line 13.
 Since the selected ${\bf G}^*_k$ may be nondeterministic which is caused by conflict decisions, for each transmitted event, its reachable (deterministic) state we choose has the maximal number of non-transmitted events (lines 17-18).    
Therefore, it can be directly implied that the obtained information transmission policy transmits fewer events than ${\bf G}^*$ by lines 13-14 and 17-18. 

\subsection{Realization of the information transmission policy} 
In the last step, we synthesize a deterministic automaton $\bf{S}$ from the largest feasible automaton ${\bf G}^*$ to present a specific information transmission policy. However, we are not able to exactly know how the information is transmitted at each step.  For instance, there exists a state $\tilde{q}\in z$ and $z'=\xi'(z,\sigma)!$ with $\tilde{q}',\tilde{q}''\in z'$ and $\tilde{q}',\tilde{q}''\in f(\tilde{q},\sigma)$, namely the states $\tilde{q}'$ and $\tilde{q}''$ are the future states of $\tilde{q}$ reached by $\sigma$ and in the same state $z'$. Hence, we cannot exactly know which state $\tilde{q}$ reaches by $\sigma$, i.e., $$|D|=|\{f(\tilde{q},\sigma)\cap z'|\neq 1.$$
What we need to clarify in Step 3 is to show which labeled state is reached at each step by the transmitted event.   
In order to clarify the transmission order of the states in $ f(\tilde{q},\sigma)\cap z'$, we employ $rank()$ defined below to reorder the states in $z'$ such that the ordered states in $rank(f(\tilde{q},\sigma)\cap z')$ satisfies that the next state can be reached by the previous state in $f(\tilde{q},\sigma)\cap z'$. Let $D=f(\tilde{q},\sigma)\cap z'=\{\tilde{q}_1,\dots,\tilde{q}_n\}\in Z'$. We define
\begin{align*}
rank(D)=\{\{\tilde{q}_{k1},\dots,\tilde{q}_{kn}\}:\ \exists s_1\dots s_{n-1}\in \Sigma^*\ s.t.\\ \tilde{q}_{ki+1}\in f(\tilde{q}_{ki},s_{i})\}.
\end{align*}
Then, we obtain that the first state in $rank(D)$ is the future state of $\tilde{q}\in z$. 
 In order to accurately know how the information is transmitted at each step, we propose Algorithm~3 in the following to synthesize a sensor automaton ${\bf A}$ and a corresponding labeling function $\mathcal{L}$, i.e. $\Omega=({\bf A},\mathcal{L})$. 
\renewcommand{\algorithmicrequire}{\textbf{Input:}}
\renewcommand{\algorithmicensure}{\textbf{Output:}}
\begin{algorithm}[h]\label{alg.4}
\caption{Synthesis of the sensor automaton ${\bf A}$ and the labeling function $\mathcal{L}$.}
 \begin{algorithmic}[1]\Require ${\bf S}=(Z',\Sigma,\xi',z_{0}')$ with  $Z'=\{z_{0},z_{1},\dots,z_{m}\}$
   \Ensure $\Omega=({\bf A},\mathcal{L})$ with ${\bf A}=(X,\Sigma,\eta, x_0)$ 
 \State  $X=W=D=\emptyset$;
\If{$|q_0\times \{Y,N\}^{|\Sigma_{q_0}|}\cap z_0|\neq 1$}
      \State $rank(z_0)=\{\tilde{q}_{01},\dots,\tilde{q}_{0d}\};$
      \State $x_0=\tilde{q}_{01};$
       \State $X=X\cup\{x_0\};$
       \State $\mathcal{L}(x_0,\sigma)=Lab_{\sigma}(x_0)$ for $\sigma\in \Sigma_{x_0}$;
         \Else
 \State $x_0=q_0\times \{Y,N\}^{|\Sigma_{q_0}|}\cap z_0$;
  \State $X=X\cup\{x_0\};$
  \State $\mathcal{L}(x_0,\sigma)=Lab_{\sigma}(x_0)$ for $\sigma\in \Sigma_{x_0}$;
  \EndIf 
 \For{ $i=0$ to $i=m$}
\For{each $\tilde{q}\in z_i$}
  \State Function $F(\tilde{q})$
\For{each $\sigma\in \Sigma$ with $\delta(q,\sigma)!$}
  \State $X=X\cup \{\tilde{q}\}$;
  \State $\mathcal{L}(\tilde{q},\sigma)=Lab_{\sigma}(\tilde{q})$;
    \begin{align*} 
    z'=
  \begin{cases}
   z,\ \mbox{if}\ Lab_{\sigma}(\tilde{q})=N\\
   \xi' (z,\sigma),\ \mbox{if}\ Lab_{\sigma}(\tilde{q})=Y;\\
   \end{cases}\\
\end{align*}
\State $D=\{f(\tilde{q},\sigma)\cap z'\}$
\State Function $DET(D)$
\State $rank(D)=\{\tilde{q}_{k1},\dots,\tilde{q}_{kh}\}$
\If{$|D|> 1$}
 \If{$\{\tilde{q}_{k1}=\eta(\tilde{q},\sigma)\}\nsubseteq W$}
       \State $X=X\cup\{\tilde{q}_{k1}\};$
       \State $W=W\cup\{\tilde{q}_{k1}=\eta(\tilde{q},\sigma)\};$
       \State $\mathcal{L}(\tilde{q}_{k1},\sigma)=Lab_{\sigma}(\tilde{q}_{k1})$;
       \State call Function $F(\tilde{q}_{k1})$
\Else
      \State $D=D\setminus \{\tilde{q}_{k1}\};$
      \State call Function $DET(D)$
\EndIf 
\EndIf 
\If{$|D|= 1$}
 \If{$\{\tilde{q}_{kh}=\eta(\tilde{q},\sigma)\}\nsubseteq W$}
        \State $X=X\cup\{\tilde{q}_{kh}\};$
       \State $W=W\cup\{\tilde{q}_{kh}=\eta(\tilde{q},\sigma)\};$
       \State call Function $F(\tilde{q}_{kh})$
\Else
\State call Function $F(\tilde{q}_{kh})$
\EndIf
\EndIf
     \EndFor
    \EndFor
  \EndFor
     \end{algorithmic}
\end{algorithm}

Algorithm 3 works as follows:
Lines 1-11 are given to ensure the initial state. The initial state may not be a singleton since there may exist states that are reached unobservable from one initial state in $q_0\times \{Y,N\}^{|\Sigma_{q_0}}$ by some string and also in $q_0\times \{Y,N\}^{|\Sigma_{q_0}}$ as well, i.e., $|q_0\times \{Y,N\}^{|\Sigma_{q_0}|}\cap z_0|\neq 1$ (line 2). We need to reorder states $q_0\times \{Y,N\}^{|\Sigma_{q_0}|}\cap z_0$ and choose the first element of $rank(q_0\times \{Y,N\}^{|\Sigma_{q_0}|}\cap z_0)$ as the initial state (lines 3-6) because the rest initial states in $q_0\times \{Y,N\}^{|\Sigma_{q_0}|}\cap z_0$ can be reached by it.
 Lines 12-43 give the transmission policy of each states at each step. Function $F(\tilde{q})$ (lines 14-41) decides the deterministic future state of $\tilde{q}$ and its transmission policy (line 25). It is recurrently called by lines 26, 36, and 38 to get the further future states. Function $DET(D)$ considers the cases that the future states of $\tilde{q}$ is unique (lines 21-31) or not (lines 32-40). If the future state is unique, then take this future state as the next state of $\tilde{q}$ (lines 34-35). Hence, the labels of the events defined at this future state give the transmission policy of these events. If the future states of $\tilde{q}$ is not unique, we need to reorder its future states (line 20) and pick the first element as the next state of $\tilde{q}$ (lines 23-25).  
We employ $W$ to record the transitions (line 24 and line 35) and ensure that the unobservable reached states in $f(\tilde{q},\sigma)\cap z'$ can get the accurate next future states by line 27.

Then, by Algorithm~3, we obtain an information transmission policy  $\Omega=({\bf A},\mathcal{L})$ which solves Problem~1 we proposed.
 An example is given in the following.

\begin{Example}
Let us  consider ${\bf G}^*$ in Fig.~\ref{fig:G3}. We synthesize an automaton $\bf{S}$ from the largest feasible automaton ${\bf G}^*$, where the obtained automaton $\bf{S}$ transmits events as few as possible by choosing the states that have the maximal number of non-transmitted events. The resulting automaton $\bf{S}$ is shown in Fig.~\ref{fig:G5}. 
Next, by Algorithm~3 we synthesize $\Omega=({\bf A},\mathcal{L})$ based on ${\bf S}$. 
\begin{enumerate}
\item[(1)]
Initially, $|q_0\times \{Y,N\}^{|\Sigma_{q_0}|}\cap z_0|= 1$, by line 8, we have that $x_0=q_0NNY$,  $X=X\cup\{q_0NNY\}=\{q_0NNY\}$, $\mathcal{L}(q_0NNY,\sigma_1)=Lab_{\sigma_1}(q_0NNY)=N$, $\mathcal{L}(q_0NNY,\sigma_2)=Lab_{\sigma_2}(q_0NNY)=N$, and $\mathcal{L}(q_0NNY,\sigma_3)=Lab_{\sigma_3}(q_0NNY)=Y$. 

Then, we need to compute the future states of the states in $z_0=\{q_0NNY,q_5,q_1Y\}$ one by one. 
\item[(2)]
For  $q_0NNY\in z_0$ and $\sigma_1\in \Sigma$ with $\delta(q_0,\sigma_1)!$, we have that $X=X\cup\{q_0NNY\}=\{q_0NNY\}$ by line 16. 
Then, we can obtain that $z'=z$ since $Lab_{\sigma_1}(q_0NNY)=N$ (by line 17). 
We have that $D=f(q_0NNY,\sigma_1)\cap z'=\{q_5\}$. 
Since $|D|=1$ and $\{q_5=\eta(q_0NNY,\sigma_1)\}\nsubseteq W$, we have that $X=X\cup\{q_5\}=\{q_0NNY,q_5\}$ and $W=\{q_5=\eta(q_0NNY,\sigma_1)\}$ by lines 30-32. 
Note that $q_5$ has no future state so that the next state in $z_0$ is considered subsequently. 
 
 \begin{figure}
  \centering
  \includegraphics[width=0.2\textwidth]{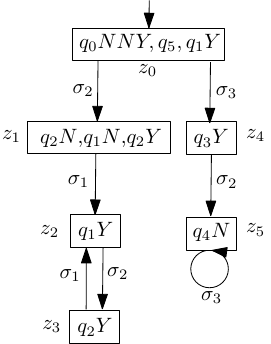}\\
  \caption{Synthesized deterministic information  transmission  policy ${\bf S}$ }
  \label{fig:G5}
\end{figure} 
 \begin{figure}
  \centering
  \includegraphics[width=0.35\textwidth]{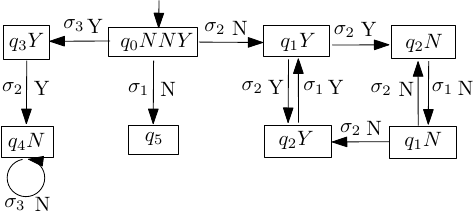}\\
  \caption{The sensor automaton $\bf{A}$}
  \label{fig:G6}
\end{figure}
\item[(3)]
For $q_0NNY\in z_0$ and $\sigma_2\in \Sigma$ with $\delta(q_0,\sigma_2)!$, we have that $z'=z$ since $Lab_{\sigma_2}(q_0NNY)=N$ by line 17.
We have that $D=f(q_0NNY,\sigma_2)\cap z'=\{q_1Y\}$. Since $|D|=1$ and $\{q_1Y=\eta(q_0NNY,\sigma_2)\}\nsubseteq W$, by lines 32-36, we have that $X=X\cup\{q_1Y\}=\{q_0NNY,q_5,q_1Y\}$, $W=W\cup\{q_1Y=\eta(q_0NNY,\sigma_2)\}$, and call Function $F(q_1Y)$.
\begin{enumerate}
    \item[(3-i)] 
For  $q_1Y\in z_0$ and $\sigma_2\in \Sigma$ with $\delta(q_1,\sigma_2)!$, we have that $X=X\cup\{q_1Y\}=\{q_0NNY,q_5,q_1Y\}$ by line 16 and $\mathcal{L}(q_1Y,\sigma_2)=Lab_{\sigma_2}(q_1Y)=Y$ by line 17. Then, we obtain that $z_1=\eta(z_0,\sigma_2)=\{q_2Y,q_2N\}$. We have that $D=f(q_1Y,\sigma_2)\cap z_1=\{q_2Y,q_2N\}$. By line 20, we have that $rank(D)=\{q_2N,q_2Y\}$. Since $|D|\neq1$ and $\{q_2N=\eta(q_1Y,\sigma_2)\}\nsubseteq W$, by lines 21-26, we have that $X=\{q_0NNY,q_5,q_1Y,q_2N\}$, $W=W\cup\{q_2N=\eta(q_1Y,\sigma_2)\}$, and call Function $F(q_2N)$.

\ \ \ \ \ \ \ \ \vdots
\end{enumerate}
\item[(4)]
For  $q_0NNY\in z_0$ and $\sigma_3\in \Sigma$ with $\delta(q_0,\sigma_3)!$, we have that $X=X\cup\{q_0NNY\}$ by line 16. Then, we obtain that $z_3=\eta(z_0,\sigma_3)=\{q_3Y\}$. We have that $D=f(q_0NNY,\sigma_3)\cap z_3=\{q_3Y\}$. Since $|D|=1$ and $Lab_{\sigma_3}(q_0NNY)=Y$ (by line 17), we have that $D=f(q_0NNY,\sigma_3)\cap z_3=\{q_3Y\}$. Since $|D|=1$ and $\{q_3Y=\eta(q_0NNY,\sigma_3)\}\nsubseteq W$, by lines 34-36, we have that $X=X\cup\{q_3Y\}$ and $W=W\cup\{q_3Y=\eta(q_0NNY,\sigma_3)\}$, and call Function $F(q_3Y)$.
\begin{enumerate}
    \item[(4-i)]
For  $q_3Y\in z_3$ and $\sigma_2\in \Sigma$ with $\delta(q_3,\sigma_2)!$, we have that $X=X\cup\{q_3Y\}$ by line 16 and $\mathcal{L}(q_3Y,\sigma_2)=Lab_{\sigma_2}(q_3Y)=Y$ by line 17. Then, we obtain that $z_4=\eta(z_3,\sigma_2)=\{q_4N\}$. We have that $D=f(q_3Y,\sigma_2)\cap z_4=\{q_4N\}$. Since $|D|=1$ and $\{q_4N=\eta(q_3Y,\sigma_2)\}\nsubseteq W$, by lines 34-36, we have that $X=X\cup\{q_4N\}$ and $W=W\cup\{q_4N=\eta(q_3Y,\sigma_2)\}$, and call Function $F(q_4N)$.

\ \ \ \ \ \  \vdots

\end{enumerate}
\end{enumerate}
The obtained $\bf{A}$ is shown in Fig.~\ref{fig:G6}.
\end{Example}

\section{Conclusion}
In this paper, we have studied the problem of optimal sensor schedules for remote state estimation of discrete-event systems. 
We have investigated a novel transmission mechanism that selectively transmits the observable events according to an information transmission policy. We have  constructed  a  dynamic  observer  that  contains  all  possible information transmission policies. Then, we have showed that  the  information  updating  rule  of  the  dynamic  observer indeed yields the state estimate of the receiver. We also have synthesized a deterministic information policy that ensures that the IS-based property is satisfied. Algorithms corresponding to the above procedures have been given.  In the future, we aim to consider the distributed and decentralized sensing and information transmission architecture based on the results proposed in this paper.

\bibliographystyle{unsrt}        
\bibliography{main}           

\begin{thebibliography}{10}

\bibitem{zhou2022state}
Yingrui Zhou, Zengqiang Chen, Zhipeng Zhang, Yuanhua Ni, and Zhongxin Liu.
\newblock A state space approach to decentralized fault se-coprognosability of
  partially observed discrete event systems.
\newblock {\em IEEE Transactions on Cybernetics}, 53(3):2028--2033, 2022.

\bibitem{ma2023verification}
Ziyue Ma, Yin Tong, and Carla Seatzu.
\newblock Verification of pattern-pattern diagnosability in partially observed
  discrete event systems.
\newblock {\em IEEE Transactions on Automatic Control}, 2023.

\bibitem{ElMaraghy2005}
Hoda~A ElMaraghy.
\newblock Flexible and reconfigurable manufacturing systems paradigms.
\newblock {\em International journal of flexible manufacturing systems},
  17:261--276, 2005.

\bibitem{hu2023design}
Yihui Hu, Ziyue Ma, and Zhiwu Li.
\newblock Design of supervisors for partially observed discrete event systems
  using quiescent information.
\newblock {\em IEEE Transactions on Automation Science and Engineering}, 2023.

\bibitem{liu2019}
Yingying Liu, Kai Cai, and Zhiwu Li.
\newblock On scalable supervisory control of multi-agent discrete-event
  systems.
\newblock {\em Automatica}, 108:108460, 2019.

\bibitem{PAIVA2021100146}
Pedro~RR Paiva, Braian~I de~Freitas, Lilian~K Carvalho, and Jo{\~a}o~C Basilio.
\newblock Online fault diagnosis for smart machines embedded in industry 4.0
  manufacturing systems: A labeled petri net-based approach.
\newblock {\em IFAC Journal of Systems and Control}, 16:100146, 2021.

\bibitem{ZHOU2020}
Yuan Zhou, Hesuan Hu, Yang Liu, Shang-Wei Lin, and Zuohua Ding.
\newblock A distributed method to avoid higher-order deadlocks in multi-robot
  systems.
\newblock {\em Automatica}, 112:108706, 2020.

\bibitem{YinRobust}
Xiang Yin, Jun Chen, Zhaojian Li, and Shaoyuan Li.
\newblock Robust fault diagnosis of stochastic discrete event systems.
\newblock {\em IEEE Transactions on Automatic Control}, 64(10):4237--4244,
  2019.

\bibitem{han2022revisiting}
Xiaoguang Han, Jinliang Wang, Zhiwu Li, Xiaoyan Chen, and Zengqiang Chen.
\newblock Revisiting state estimation and weak detectability of discrete-event
  systems.
\newblock {\em IEEE Transactions on Automation Science and Engineering},
  20(1):662--674, 2022.

\bibitem{li2022synthesis}
Jinglun Li and Shigemasa Takai.
\newblock Synthesis of maximally permissive supervisors for similarity control
  of partially observed nondeterministic discrete event systems.
\newblock {\em Automatica}, 135:109978, 2022.

\bibitem{ALIKHANI2021109575}
Hamid Alikhani and Nader Meskin.
\newblock Event-triggered robust fault diagnosis and control of linear roesser
  systems: A unified framework.
\newblock {\em Automatica}, 128:109575, 2021.

\bibitem{li2023error}
Yuting Li, Christoforos~N Hadjicostis, Naiqi Wu, and Zhiwu Li.
\newblock Error-and tamper-tolerant state estimation for discrete event systems
  under cost constraints.
\newblock {\em IEEE Transactions on Automatic Control}, 2023.

\bibitem{yang2022secure}
Shuo Yang and Xiang Yin.
\newblock Secure your intention: On notions of pre-opacity in discrete-event
  systems.
\newblock {\em IEEE Transactions on Automatic Control}, 2022.

\bibitem{alves2023state}
Marcos~VS Alves and Jo{\~a}o~C Basilio.
\newblock State estimation and detectability of networked discrete event
  systems with multi-channel communication networks.
\newblock {\em IEEE Transactions on Automation Science and Engineering}, 2023.

\bibitem{liu2021improved}
Yang Liu, Zhaocong Liu, Xiang Yin, and Shaoyuan Li.
\newblock An improved approach for verifying delayed detectability of
  discrete-event systems.
\newblock {\em Automatica}, 124:109291, 2021.

\bibitem{hou2023modeling}
Yunfeng Hou, Yunfeng Ji, Gang Wang, Ching-Yen Weng, and Qingdu Li.
\newblock Modeling and state estimation for supervisory control of networked
  timed discrete-event systems and their application in supervisor synthesis.
\newblock {\em International Journal of Control}, pages 1--21, 2023.

\bibitem{yin2016uniform}
Xiang Yin and St{\'e}phane Lafortune.
\newblock A uniform approach for synthesizing property-enforcing supervisors
  for partially-observed discrete-event systems.
\newblock {\em IEEE Transactions on Automatic Control}, 61(8), 2016.

\bibitem{yin2019synthesis}
Xiang Yin and Shaoyuan Li.
\newblock Synthesis of dynamic masks for infinite-step opacity.
\newblock {\em IEEE Transactions on Automatic Control}, 65(4):1429--1441, 2019.

\bibitem{yin2018minimization}
Xiang Yin and St{\'e}phane Lafortune.
\newblock Minimization of sensor activation in decentralized discrete-event
  systems.
\newblock {\em IEEE Transactions on Automatic Control}, 63(11):3705--3718,
  2018.

\bibitem{zhang2015maximum}
Bo~Zhang, Shaolong Shu, and Feng Lin.
\newblock Maximum information release while ensuring opacity in discrete event
  systems.
\newblock {\em IEEE Transactions on Automation Science and Engineering},
  12(3):1067--1079, 2015.

\bibitem{rudie2003minimal}
Karen Rudie, St{\'e}phane Lafortune, and Feng Lin.
\newblock Minimal communication in a distributed discrete-event system.
\newblock {\em IEEE Transactions on Automatic Control}, 48(6):957--975, 2003.

\bibitem{liu2022}
Yingying Liu, Xiang Yin, and Shaoyuan Li.
\newblock To transmit or not to transmit: Optimal sensor scheduling for remote
  state estimation of discrete-event systems.
\newblock In {\em 2022 American Control Conference (ACC)}, pages 4483--4489.
  IEEE, 2022.

\end{thebibliography}



\end{document}